\documentclass[preprint,showpacs,aps]{revtex4}
\usepackage{amsmath}

\input{tcilatex}

\begin{document}

\title{Compactified Newtonian Potential and a Possible Explanation for Dark
Matter}
\author{Hongya Liu}
\email{hyliu@dlut.edu.cn}
\affiliation{Department of Physics, Dalian University of Technology, Dalian 116024, P.R.
China}

\begin{abstract}
Exact Newtonian potential with an extra space dimension compactified on a
circle is derived and studied. It is found that a point mass located at one
side of the circle can generate almost the same strength of gravitational
force at the other side.\ Combined with the brane-world scenarios, it means
that although one can not see a particle which is confined to another brane,
one can feel its gravity explicitly as if it were located on our brane. This
leads to the conclusion that matter on all branes may contribute equally to
the curvature of our brane. Therefore, dark matter is probably nothing
unusual but a `shadow' cast on our brane by hidden ordinary matter of other
branes. It is shown that the physical effect of this `shadow' could be
modeled by a perfect fluid with an effective mass density and a non-zero
pressure. This fluid could serve as a dark matter candidate with an equation
of state $p_{eff}=\omega \rho _{eff}$ where $\omega $ is a positive $r$%
-dependent function and $\omega \neq 0,1/3$,\ implying that it is neither
cold nor hot. So, if the higher-dimensional universe contains several
branes, the total amount of hidden matter would be able to provide the
required amount for dark matter. Meanwhile, dark matter halos of galaxies
might be just halos of the effective mass distribution yielded by hidden
ordinary stars, and these stars are confined on other branes but captured by
our galaxies. Some more properties and implications are also discussed.
\end{abstract}

\maketitle

The idea of extra compact spatial dimensions was firstly used in
Kaluza-Klein theory \cite{Kaluza} to unify gravity with electromagnetism and
then developed in superstrings, supergravity, and M-theories to unify all
the four fundamental forces in nature. It was firstly pointed out by
Arkani-Hamed, Dimopoulos and Dvali (ADD) that the well known Newtonian
inverse square law could be violated if there exist extra space dimensions.
They showed that if the $N$ extra dimensions are curled up with radius $R$,
the Newtonian force between two point masses\ would undergo a transition
from the familiar form $1/r^{2}$ at large distances ($r\gg R$) to a form $%
1/r^{2+N}$\ at small distances ($r\ll R$) \cite{ADD98}\cite{ADD99}. Thus the 
$(4+N)$ dimensional gravity and other interactions could be united at the
electroweak scale and the hierarchy problem could be resolved. This has
attracted much attention in recent years from both theoretical and
experimental physicists. Nowadays, a dozen groups over the world have set up
to look for departure from the inverse square law at small distances \cite%
{Experi}. Therefore, it is of great interest to know how the force varies
along the $r$-direction from $1/r^{2}$ to $1/r^{2+N}$ in more details.

It was shown \cite{Flor99}\cite{Kehag00} that as the separation $r$
decreases, a Yukawa term appears as a first correction to the Newtonian $1/r$
potential,%
\begin{equation}
V_{3+N}(r)\approx -\frac{G_{4}M}{r}\left( 1+\alpha e^{-r/\lambda }\right) \;,
\label{Yukawa}
\end{equation}%
where $\alpha =2N$ for\ the compactification with $N$-torus, and $\alpha =N+1
$ for an $N$-sphere. However, this approximate expression can not be used at
small separation with $r\sim \lambda $ at which the deviation might be very
large. So a global exact solution for the Newtonian potential is of
particular importance in both the experimental calculation and theoretical
analysis. In what follows, we study the simplest case with $N=1$. Here we
should mention that for $N=1$ the Yukawa parameters in (\ref{Yukawa}) are $%
\alpha =2$ and $\lambda =R$. Using the $\alpha -\lambda $ plot of Ref. \cite%
{Experi}, we find that experimental constraint on the $N=1$ case is $R<0.4mm$%
. Therefore, $N=1$ is not experimentally excluded at all; it is just
excluded by the ADD unification for the purpose to resolve the hierarchy
problem.

The ($3+1$) dimensional Newtonian potential $V_{3+1}$\ satisfies the ($3+1$)
dimensional Poisson equation%
\begin{equation}
\nabla _{3+1}^{2}V_{3+1}=2\pi ^{2}G_{4+1}\sigma   \label{Poisson}
\end{equation}%
where $G_{4+1}$ is the Newtonian constant in ($4+1$) dimensional gravity.
Consider a point mass $M$ located at the origin ($r=y=0$) and suppose the
extra dimension $y$ is compactified to a circle with radius $R$ and
circumference $L=2\pi R$. To mimic a compact dimension, we follow ADD \cite%
{ADD98}\cite{ADD99} to use a non-compact dimension with ``mirror'' masses
placed periodically along the $y$-axis. Thus we obtain%
\begin{equation}
V_{3+1}=-\frac{G_{4+1}M}{2}\sum\limits_{n=-\infty }^{\infty }\frac{1}{%
r^{2}+\left( nL-y\right) ^{2}}\;,  \label{phiSum}
\end{equation}%
which satisfies the periodic identification $y\rightarrow L\pm y$. Then we
can show that%
\begin{equation}
\frac{1}{r^{2}+\left( nL-y\right) ^{2}}=\frac{1}{2iLr}\left( \frac{1}{n+z}-%
\frac{1}{n+z^{\ast }}\right) ,\qquad z=\frac{-y-ir}{L}\;.  \label{n+Z}
\end{equation}%
Using this relation in (\ref{phiSum}), we bring the Newtonian potential $%
V_{3+1}$ to a meromorphic function. Then, with use of the identity 
\begin{equation}
\sum\limits_{n=-\infty }^{\infty }\frac{1}{z+n}=\pi \cot \pi z\;
\label{meromorphic}
\end{equation}%
in (\ref{phiSum}), we finally obtain%
\begin{equation}
V_{3+1}=-\frac{G_{4+1}M}{4Rr}\left( \frac{\sinh \frac{r}{R}}{\cosh \frac{r}{R%
}-\cos \frac{y}{R}}\right) \;.  \label{phiAnal}
\end{equation}%
This is the global exact Newtonian potential for $N=1$ which was firstly
derived by Floratos and Leontaris with use of the method of Fourier
transform \cite{Flor99}. Here we derived it directly from the ADD model (\ref%
{phiSum}). The periodicity for $y$ along the circle can be seen clearly from
(\ref{phiAnal}). One can check it by directly substituting (\ref{phiAnal})
into the ($3+1$) dimensional Laplace equation%
\begin{equation}
\left( \frac{\partial ^{2}}{\partial r^{2}}+\frac{2}{r}\frac{\partial }{%
\partial r}+\frac{\partial ^{2}}{\partial y^{2}}\right) V_{3+1}=0\;,
\label{Laplas}
\end{equation}%
which is found to be satisfied by (\ref{phiAnal}) as expected. Thus we can
define the Newtonian force along the $r$-direction as%
\begin{equation}
F_{3+1}\equiv -\frac{\partial }{\partial r}V_{3+1}=-\frac{G_{4+1}M}{4Rr^{2}}%
\left[ \frac{\sinh \frac{r}{R}}{\cosh \frac{r}{R}-\cos \frac{y}{R}}-\frac{r}{%
R}\frac{1-\cosh \frac{r}{R}\cos \frac{y}{R}}{\left( \cosh \frac{r}{R}-\cos 
\frac{y}{R}\right) ^{2}}\right] \;.  \label{r-force}
\end{equation}

Equations (\ref{phiAnal}) and (\ref{r-force}) are exact global expressions
valid everywhere except at the origin ($r=y=0$) at which the point mass $M$
is located. At large distances $r\gg L$, these two equations lead to%
\begin{equation}
V_{3+1}=-\frac{G_{4+1}M}{4Rr}\left( 1+2e^{-\frac{r}{R}}\cos \frac{y}{R}%
+...\right) ,\;\text{for }r\gg R\;,  \label{phi-r>>L}
\end{equation}%
and%
\begin{equation}
F_{3+1}\equiv -\frac{G_{4+1}M}{4Rr^{2}}\left[ 1+2\left( 1+\frac{r}{R}\right)
e^{-\frac{r}{R}}\cos \frac{y}{R}+...\right] ,\;\text{for }r\gg R\;.
\label{F-r>>L}
\end{equation}%
At small distances $r,\left| y\right| \ll R$, we let $r/R\sim y/R\sim
\varepsilon $ with $\varepsilon $ being a small parameter, then the two
equations (\ref{phiAnal}) and (\ref{r-force}) give%
\begin{equation}
V_{3+1}=-\frac{G_{4+1}M}{2\left( r^{2}+y^{2}\right) }\left[ 1+\frac{%
r^{2}+y^{2}}{12R^{2}}+O(\varepsilon ^{4})\right] ,\;\text{for }r,\left|
y\right| \ll R\;,  \label{V-<<R}
\end{equation}%
and%
\begin{equation}
F_{3+1}=-\frac{G_{4+1}Mr}{\left( r^{2}+y^{2}\right) ^{2}}\left[
1+O(\varepsilon ^{4})\right] ,\qquad \text{for\ }r,\left| y\right| \ll R\;.
\label{F-<<R}
\end{equation}%
Thus the 4-dimensional Newtonian constant $G_{4}$ is found to be%
\begin{equation}
G_{4}=\frac{G_{4+1}}{4R}\;  \label{G_4}
\end{equation}%
as is given by ADD \cite{ADD98}\cite{ADD99}.

In the brane-world scenarios \cite{Witten}\cite{ADD98}\cite{ADD99}, our
universe is a 3-brane embedded in a higher dimensional space. While gravity
can freely propagate in all dimensions, the standard matter particles and
forces are confined to the 3-brane only. For $N=1$, it is natural to have
two 3-branes located at $y=0$ and $y=\pi R$, respectively. Suppose we live
in the visible $y=0$ brane. Then the Newtonian potential and force for a
visible point mass $M$ located at $r=y=0$ are, respectively,%
\begin{eqnarray}
V_{vis} &=&-\frac{G_{4}M}{r}\left( \frac{\sinh \frac{r}{R}}{\cosh \frac{r}{R}%
-1}\right) \;,  \notag \\
V_{vis} &=&-\frac{G_{4}M}{r}\left( 1+2e^{-\frac{r}{R}}+...\right) ,\;\text{%
for }r\gg R\;.\;  \label{V-vis}
\end{eqnarray}%
and%
\begin{eqnarray}
F_{vis} &\equiv &-\frac{\partial }{\partial r}V_{vis}=-\frac{G_{4}M}{r^{2}}%
\left( \frac{\sinh \frac{r}{R}+\frac{r}{R}}{\cosh \frac{r}{R}-1}\right) \;, 
\notag \\
F_{vis} &\equiv &-\frac{G_{4}M}{r^{2}}\left[ 1+2\left( 1+\frac{r}{R}\right)
e^{-\frac{r}{R}}+...\right] ,\;\text{for }r\gg R\;,  \notag \\
F_{vis} &\rightarrow &-\frac{G_{4+1}M}{r^{3}}\;,\qquad \text{for\ }%
r\rightarrow 0\;.  \label{F-vis}
\end{eqnarray}

Another case that a point mass $M$ is confined to the hidden side of the
circle ($r=0,$ $y=\pi R$) might be more interesting. For this case we should
replace $y/R$ with $\left( \pi +y/R\right) $ in the general solution (\ref%
{phiAnal}), then the Newtonian potential and force for a hidden point mass $%
M $ located at $r=0$, $y=\pi R$\ are%
\begin{eqnarray}
V_{hid} &=&-\frac{G_{4}M}{r}\left( \frac{\sinh \frac{r}{R}}{\cosh \frac{r}{R}%
+1}\right) \;,  \notag \\
V_{hid} &=&-\frac{G_{4}M}{r}\left( 1-2e^{-\frac{r}{R}}+...\right) ,\;\text{%
for }r\gg R\;.\;  \label{V-hid}
\end{eqnarray}%
and%
\begin{eqnarray}
F_{hid} &\equiv &-\frac{\partial }{\partial r}V_{hid}=-\frac{G_{4}M}{r^{2}}%
\left( \frac{\sinh \frac{r}{R}-\frac{r}{R}}{\cosh \frac{r}{R}+1}\right) \;, 
\notag \\
F_{hid} &\equiv &-\frac{G_{4}M}{r^{2}}\left[ 1-2\left( 1+\frac{r}{R}\right)
e^{-\frac{r}{R}}+...\right] ,\;\text{for }r\gg R\;,  \notag \\
F_{hid} &\rightarrow &-\frac{G_{4+1}M}{12R^{3}}r\rightarrow 0\;,\qquad \text{%
for\ }r\rightarrow 0\;.  \label{F-hid}
\end{eqnarray}%
Thus we see that $V_{hid}$\ and $F_{hid}$\ are regular everywhere including
at $r=0$. So one can not see the particle itself because it is hidden in
another side of the extra dimension. However, one can ``feel'' its
gravitational force, and this force is, to the leading term at large
separation, of the same strength as if that particle $M$ is located at our
side. Comparing (\ref{V-vis}), (\ref{V-hid}) with (\ref{Yukawa}), we also
find that a difference between these two cases appears in the first
correction Yukawa term: If the source particle is at our side, the Yukawa
parameter is $\alpha =2$; if it is at the hidden side, it is $\alpha =-2$.

Equations (\ref{V-vis}) and (\ref{V-hid}) enable us to use the superposition
theorem to write down the Newtonian potential for discrete particles,%
\begin{equation}
V(r)=-G_{4}\left[ \sum_{i}\frac{M_{vis}^{i}}{\left| \mathbf{r-r}_{i}\right| }%
\left( \frac{\sinh \frac{\left| \mathbf{r-r}_{i}\right| }{R}}{\cosh \frac{%
\left| \mathbf{r-r}_{i}\right| }{R}-1}\right) +\sum_{i}\frac{M_{hid}^{i}}{%
\left| \mathbf{r-r}_{i}\right| }\left( \frac{\sinh \frac{\left| \mathbf{r-r}%
_{i}\right| }{R}}{\cosh \frac{\left| \mathbf{r-r}_{i}\right| }{R}+1}\right) %
\right] ,  \label{V-suppos}
\end{equation}%
where $M_{vis}$\ and $M_{hid}$\ are masses of particles located at the
visible and hidden brane, respectively. At large distances, i.e., $\left| 
\mathbf{r-r}_{i}\right| \gg R$ for all these particles, equation (\ref%
{V-suppos}) approaches to%
\begin{equation}
V(r)\approx -G_{4}\left( \sum_{i}\frac{M_{vis}^{i}}{\left| \mathbf{r-r}%
_{i}\right| }+\sum_{i}\frac{M_{hid}^{i}}{\left| \mathbf{r-r}_{i}\right| }%
\right) \;,\qquad \text{for }\left| \mathbf{r-r}_{i}\right| \gg R\;.\text{\ }
\label{V-supAp}
\end{equation}%
Similar results can be written down if matter is continuously distributed
over the branes. Thus, from (\ref{V-supAp}), we conclude that, \textit{at
large distances, particles at the hidden side of the extra dimension
contribute to the Newtonian potential with the same strength as if they were
at the visible side}.

This is an important result which implies that all particles, no matter they
are located at our brane or not, will \textit{equally} contribute to the
curvature of our brane. ADD and other authors have already discussed this
case in more detail and for a variety of\ brane models \cite{ADD99}\cite%
{Manyfold00}. Here, in this paper, we provide with an exactly solvable model
to exhibit it explicitly as given above.

Let us discuss it from a purely three-dimensional point of view. In (\ref%
{V-hid}), although the source particle $M$ is hidden at another side, it
generates, in our 3D surface, a Newtonian potential from which we can define
an effective mass density $\rho _{eff}$ via%
\begin{equation}
\rho _{eff}\equiv \frac{1}{4\pi G_{4}}\nabla _{3}^{2}V_{hid}=\frac{M}{4\pi
R^{2}}\frac{\sinh \frac{r}{R}}{r\left( \cosh \frac{r}{R}+1\right) ^{2}}\;.
\label{rhoEff}
\end{equation}%
This density is regular everywhere including at $r=0$, at which it reaches
to an finite maximum $M(16\pi R^{3})^{-1}$.\ Using (\ref{rhoEff}) we can
calculate the total amount of effective mass inside a shell of radius $r$,
which is%
\begin{equation}
\mu (r)\equiv 4\pi \int_{0}^{r}\rho _{eff}r^{2}dr=M\frac{\sinh \frac{r}{R}-%
\frac{r}{R}}{\cosh \frac{r}{R}+1}\;.  \label{mu(r)}
\end{equation}%
We find $\mu (r)\rightarrow M$ as $r/R\rightarrow \infty $. So \textit{the
total amount of the effective mass distributed over our brane equals exactly
to the same mass of the hidden particle}; it does not depend on the size $R$
of the extra dimension at all, though the density $\rho _{eff}$ does depend
on $R$.

Thus we arrive at a conclusion that hidden particles contribute to our brane
with the same amount of effective mass. This conclusion is drawn from the
special case with only one extra dimension, however, it is reasonable to
expect it to hold for more extra compact dimensions as discussed by ADD in
Ref. [3]. So, as a whole, all matter hidden on other branes may yield an
effective mass density distributed over our brane. This kind of mass density
is ``dark'' because one can not ``see'' it. But its gravity contributes to
the curvature of our brane and can be measured without any problem. Thus the
effective mass distribution provides us with a natural candidate for dark
matter. Be aware that the amount of dark matter required by observations is
about six times of the visible ordinary matter \cite{Spergel03}. Meanwhile,
the string-membrane theories require up to seven extra dimensions. So if
there are several branes and each brane has the same order of amount of
ordinary matter as ours, then the required amount of dark matter could be
found. If we accept this explanation, then \textit{dark matter probably is
nothing unusual but a `shadow' cast on our brane by\ hidden ordinary matter
of other branes}.

Now let us go further to study the effective mass distribution $\rho _{eff}$
in Eq. (\ref{rhoEff}). If we explain $\rho _{eff}$ as a dark matter
candidate, then this kind of dark matter has a significant property: As a
whole, it yields an attractive force to other particles as seen from (\ref%
{F-hid}), --- this can explain the dark halos of galaxies as well as the
mass density for the universe. However, $\rho _{eff}$ is spread\ over our
brane implying that different parts of itself can not collapse. This is
distinctive from other dark matter candidates but does not contradict with
current observations. To explain it, a natural way is to introduce an
effective pressure $p_{eff}$. So we obtain a perfect fluid model for which
we can write the hydrostatic equilibrium equation as%
\begin{equation}
\frac{dp_{eff}}{dr}=-\frac{G_{4}\mu (r)\rho _{eff}}{r^{2}}\;.  \label{dp/dr}
\end{equation}%
Substituting (\ref{mu(r)}) and (\ref{rhoEff}) in this equation, we obtain%
\begin{equation}
p_{eff}=\frac{G_{4}M^{2}}{4\pi R^{2}}\int\nolimits_{r}^{\infty }\frac{\sinh 
\frac{r}{R}\left( \sinh \frac{r}{R}-\frac{r}{R}\right) }{r^{3}\left( \cosh 
\frac{r}{R}+1\right) ^{3}}dr  \label{p-eff-int}
\end{equation}%
where we have chosen $p_{eff}$ such that $p_{eff}\rightarrow 0$ as $%
r\rightarrow \infty $. This corresponds to an equation of state $%
p_{eff}=\omega \rho _{eff}$ with $\omega $ being a function of the
coordinate $r$, i.e., $\omega =\omega (r)$. We can show that $\omega $ is
positive but not equal to zero or\ $1/3$. \textit{So this kind of dark
matter is neither cold (}$\omega =0$\textit{) nor hot (}$\omega =1/3$\textit{%
)}.

In 5D relativity, the Newtonian potential and Newtonian force derived in
this paper should be recovered as the corresponding Newtonian limit. For the
exterior field of a point mass, the field equations should be the vacuum
ones, $R_{ab}=0$ ($a,b=0,1,2,3;5$). It is known from the induced matter
theory \cite{Wesson} \cite{Overduin} that five-dimensional Ricci-flat
equations $R_{ab}=0$ contain the four-dimensional Einstein equations, which
in terms of the 4D Einstein tensor and an induced energy-momentum tensor are%
\begin{equation}
^{(4)}G_{\alpha \beta }=\kappa T_{\alpha \beta }^{eff}\;.  \label{4DEinEqs}
\end{equation}%
It is also known that this induced energy-momentum tensor $T_{\alpha \beta
}^{eff}$ could be modeled by a perfect fluid with an effective mass density
and an effective pressure for the 5D spherically symmetric static solutions %
\cite{LiuJMP92}, the 5D cosmological solutions \cite{PoncedeLeon88} \cite%
{LiuMash95} \cite{LiuWessonAPJ} and some other kinds of solutions \cite%
{Wesson} \cite{Overduin}. This is guaranteed by Campbell's theorem which
states that any solution of the Einstein equations in $N$-dimensions can be
locally embedded in a Ricci-flat manifold of ($N+1$)-dimensions \cite%
{Campbell}. Thus we conclude that the 4D effective energy-momentum tensor $%
T_{\alpha \beta }^{eff}$ appeared in (\ref{4DEinEqs}) is actually the
counterpart of the effective $\rho _{eff}$ and $p_{eff}$ of (\ref{rhoEff})
and (\ref{p-eff-int}) of the Newtonian model. We also conclude that the
induced matter theory could occupy a central position in the description of
dark matter in brane world scenarios.

From equations (\ref{V-supAp}) and (\ref{V-suppos}) we see clearly that a
hidden particle located at another brane can generate almost the same
strength of force on our brane. So, generally speaking, a visible star of
our brane could be captured by a hidden star of another brane and thus forms
a special kind of binary. A visible galaxy of our brane can capture hidden
stars from other branes, and this would provide us with a possible
interpretation for the dark halos of galaxies. A hidden galaxy of another
brane can also capture visible matter from our brane. Recently, it was
reported \cite{NewSci} that Simon, Robishaw and Blitz observed a cloud of
hydrogen gas rotating fast around a devoid region and they called this
region as a dark galaxy. If it is true, this dark galaxy could be just the 
\textit{`shadow'} of a galaxy which is hidden on another brane. We should
emphasize that all these speculations are based on the possible existence of
extra compact space dimensions which is required by Kaluza-Klein and by
string-membrane theories. We hope that future observations may show whether
these speculations are correct or not and whether there exist extra
dimensions. From our 4D point of view, the net effect of matter hidden on
all other branes could be described by an effective mass distribution or an
effective energy-momentum tensor which could serve as a candidate for dark
matter. This candidate may provide us with the same mathematical results as
provided by other dark matter candidates, with an exception that the
physical nature of this dark matter could be just the higher dimensional
gravitational field or gravitons.

\begin{acknowledgments}
We thank James Overduin for comments. This work was supported by National
Natural Science Foundation (10273004) and National Basic Research Program
(2003CB716300) of P. R. China.
\end{acknowledgments}

\end{document}